\documentclass[dvips]{article}
\usepackage{icrctc07}

\newcommand{\QSi}{Q_{\mbox{\scriptsize Si}}}
\newcommand{\QSci}{Q_{\mbox{\scriptsize Sci}}}

\title{Relative abundances of cosmic ray nuclei B-C-N-O in the energy
  region from 10 GeV/n to 300 GeV/n. Results from ATIC-2 (the science
  flight of ATIC).}
\shorttitle{Relative abundances}
\authors{%
A. D. Panov$^{1}$, 
N. V. Sokolskaya$^{1}$,
J. H. Adams, Jr.$^{2}$, 
H. S. Ahn$^{3}$, 
G. L. Bashindzhagyan$^{1}$,
K. E. Batkov$^{1}$,
J. Chang$^{4,5}$, 
M. Christl$^{2}$, 
A. R. Fazely$^{6}$,
O. Ganel$^{3}$,
R. M. Gunasingha$^{6}$,
T. G. Guzik$^{7}$,
J. Isbert$^{7}$,
K. C. Kim$^{3}$,
E. N. Kouznetsov$^{1}$,
M. I. Panasyuk$^{1}$,
W. K. H. Schmidt$^{5}$,
E. S. Seo$^{3}$,
J. Watts$^{2}$,
J. P. Wefel$^{7}$, 
J. Wu$^{3}$,
V. I. Zatsepin$^{1}$.} 
\shortauthors{A. D. Panov and et al}
\afiliations{%
$^1$Skobeltsyn Institute of Nuclear Physics, Moscow State University, Moscow, Russia
$^2$Marshall Space Flight Center, Huntsville, AL, USA
$^3$University of Maryland, Institute for Physical Science \& Technology, College Park, MD, USA
$^4$Purple Mountain Observatory, Chinese Academy of Sciences, China
$^5$Max-Planck Institut for Solar System Research, Katlenburg-Lindau, Germany
$^6$Southern University, Department of Physics, Baton Rouge, LA, USA
$^7$Louisiana State University, Department of Physics and Astronomy, Baton Rouge, LA, USA}

\email{panov@dec1.sinp.msu.ru}

\abstract{The ATIC balloon-borne experiment measures the energy
  spectra of elements from H to Fe in primary cosmic rays from about
  100 GeV to 100 TeV.  ATIC is comprised of a fully active bismuth
  germanate calorimeter, a carbon target with embedded scintillator
  hodoscopes, and a silicon matrix that is used as the main charge
  detector. The silicon matrix produces good charge resolution for 
  protons and helium but only partial resolution for heavier nuclei.
  In the present paper, the charge resolution of ATIC was
  improved and backgrounds were reduced in the region from
  Be to Si by using the upper layer of the scintillator hodoscope
  as an additional charge detector. The flux ratios of nuclei B/C, C/O, 
  N/O in the energy region from about 10 GeV/nucleon to 300 GeV/nucleon
  obtained from this high-resolution, high-quality charge spectra are 
  presented, and compared with existing theoretical predictions.}

\begin{document}
\maketitle

\section{Introduction}

The ATIC spectrometer, its calibration and the
algorithm of trajectory reconstruction have been
described \cite{ATIC_GUZIC2004,ATIC_MSU2004A,ATIC_SOKOLSKAYA2005}. 
Charge resolution provided by the silicon matrix is sufficient to
obtain spectra of primary protons and helium
\cite{ATIC_WEFEL_PUNE2005,ATIC_PANOV2006} and preliminary spectra of
some abundant heavy nuclei \cite{ATIC_PANOV2004,ATIC_PANOV2006}. 

Very important to understand the mechanism of propagation
of cosmic rays in the Galaxy is the boron (which is a secondary nuclide) to carbon ratio in cosmic rays. The problem
of B/C ratio has been experimentally investigated in the
energy range 0.5--50~GeV/n (see \cite{HEAO_ENGELMANN1990} and
references herein).  The energy range of the ATIC experiment allow data for higher energies (up to 200--300~GeV/n) to be obtained. But
there are obstacles: 1) low charge resolution
of the silicon matrix in the range of charges 5-8 and 2) high
background in the silicon matrix charge spectrum in the range of
boron and carbon (see fig.~\ref{fig1}). In this paper we use the upper
layer of the scintillator hodoscope to improve the charge resolution 
and to reduce backgrounds in B-C region to measure B/C
in the ATIC experiment.  

\section{Improved charge spectrum}

\begin{figure}[t]
\begin{center}
\includegraphics [width=0.48\textwidth]{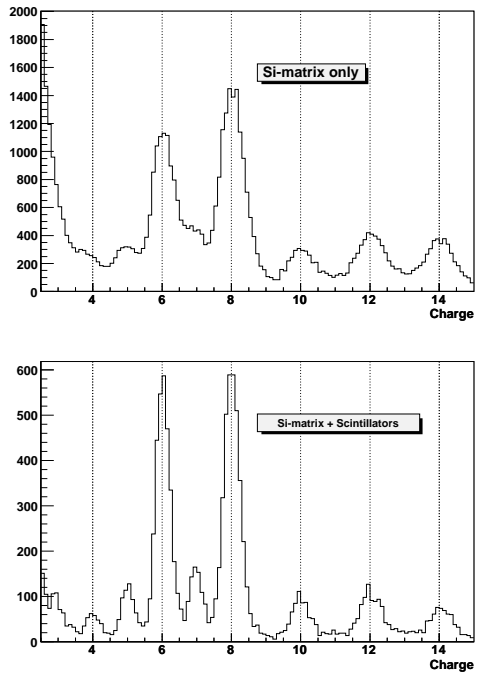}
\end{center}
\caption{The charge spectra obtained with the silicon matrix
  only and with the silicon matrix plus the upper layer of hodoscope
  for the range of energy deposit in BGO calorimeter 50--100~GeV
  (primary energy per particle approximately
  150--300~GeV)}\label{fig1}
\end{figure}
The upper scintillator layer of the hodoscope is comprised by 42
parallel scintillator strips $1\times 2\times 88.2$~cm$^3$.  
Using these scintillators as a supplementary charge
detector, involves a multi-step procedure of calibration and normalization of
the signals which will be described in detail
elsewhere. In brief, the method is the following.
The first step is to use the usual method to measure the charge
of primary particles -- trajectory reconstruction from signals in
BGO-calorimeter project - to the silicon matrix. The charge detected in the silicon matrix
$(\QSi)$ is the maximal signal in the area of confusion for the
trajectory 
\cite{ATIC_GUZIC2004,ATIC_MSU2004A,ATIC_SOKOLSKAYA2005}. In the second
step, we find the charge detected in each scintillator strip and
select the strip with the charge nearest to the charge detected by the
silicon matrix $(\QSci)$. This charge is accepted if the distance
from the strip to the reconstructed trajectory is less than 5cm and is
rejected otherwise.  In the third step, an event is rejected if $|\QSci -
\QSi| > 0.25$. The final result for the charge is $Q =
(\QSci + \QSi) / 2$.  This procedure reduces the backgrounds
in the charge spectrum and improves its resolution, but reduces the
initial statistics by a factor of about 4. There are other
strategies to process charge data from the scintillator hodoscope, but
for this paper we select the strategy of ``higher resolution - lower
backgrounds - lower statistics''. The charge spectra obtained with the
silicon matrix only and with the silicon matrix plus the upper layer
of hodoscope are compared in fig.~\ref{fig1}. 

\section{Measurement of the relative fluxes}

We calculate the ratio of fluxes of different nuclei in cosmic rays to the
flux of carbon against energy of particles per nucleon. It is a
multi-step procedure which is designed to obtain the most exact
information for fluxes of nuclei with charges $4 \le q \le 14$. 

1. For five ranges of the energy deposit $E_d$ in the BGO calorimeter
(50--100, 100--200, 200--398, 398--794, 794--1585~GeV) we obtain the charge
spectra (similar to fig.~\ref{fig1}, lower graph), and 
decompose each by
Gaussian fits (the value $\chi^2$ per degree of freedom is close
to 1 in all cases). The positions of peaks are determined and 
charge cuts are developed for each particular primary particle such that the
margins of cuts are at the half of path between adjacent peaks. The
number of counts $I^0_{s,q}$ in each charge bin $q$ for $E_d$ range
number $s$ is the raw data to obtain the fluxes of primary particles
($s=0$ corresponds to the energy region of $E_d$ 50--100~GeV, etc). 

2. Protons and helium interacting in the material (aluminum
honeycomb and other) of ATIC above the silicon matrix can
sometimes simulate heavier nuclei. This effect is energy dependent
(grows with energy).  Corresponding backgrounds $B^{\mbox{\scriptsize
    p,He}}_{s,q}$ for each value $I^0_{s,q}$ are calculated by
simulation of propagation of protons and helium through the ATIC
instrument by the FLUKA code \cite{FASSO2003}, with simulation of the
conditions of charge selection (see previous section). The apparatus
charge line widths are accounted for as well.  This procedure produces
the corrected values of intensities $I^1_{s,q} = I^0_{s,q} -
B^{\mbox{\scriptsize p,He}}_{s,q}$. The value of $B^{\mbox{\scriptsize
    p,He}}_{s,q}$ for boron $(q=5)$ varies from 9\% to 36\% of
$I^0_{s,q}$. 

3. Particles with charges $q \ge 15$ fragmenting in the material
above the silicon matrix can also produce
nuclei of $4 \le q \le 14$. The corresponding backgrounds were subtracted
but the effect is small (about 0.1\% for boron) and we do not describe
the method of subtraction here. 

4. Each nuclei of $4 \le q \le 14$ due to interactions in ATIC,
and due to the apparatus broadening of the peaks, produces a
``charge response'' of the device which may be described for each
particular nucleus $q$ at the entrance to the instrument by a set of the
coefficients $K^q_4,K^q_5,\dots,K^q_{14}$, where $K^q_4$ is the
probability to find nucleus $q$ in the charge bin $4$, etc. Of course
the coefficient $K^q_q$ dominates strongly in the set
$K^q_4,\dots,K^q_{14}$. In other words the matrix $||K^q_{q'}||$ is
diagonally-dominated. Let $F_q$ be the intensity of the nucleus $q$ at
the entrance to ATIC. Then, for each energy region $s$, the
experimental charge spectrum $I^1_q$ after subtraction of p-He
backgrounds (described above; here, we do not write the index
$s$ for simplicity) may be written as
\begin{equation}
  \label{eq:F}
  \begin{array}{l}
    I^1_4 = K^4_4 F_4 + K^5_4 F_5 +\dots+K^{14}_4F_{14}\\
    \dots\dots\dots\dots\dots\dots\dots\dots\dots\dots\dots\\
    I^1_{14} = K^4_{14} F_4 + K^5_{14} F_5 + \dots +K^{14}_{14} F_{14}
  \end{array}
\end{equation}
Eq.~(\ref{eq:F}) is a square linear system relative to unknown values
$F_4, F_5,\dots,F_{14}$ with diagonally-dominated matrix elements, and it can be
easily solved by usual methods. The coefficients of the system
$K^q_{q'}$ are calculated by simulation of propagation of different
nuclei through ATIC with FLUKA and the values $I^1_q$ are
already known after steps 2 and 3. The ratio of B/C (calculated as the
ratio of the contents of the related charge bins) reduced by
14\%--42\% for different energies. 

5. The next step is a transition from the spectra of fluxes 
as a functions of $E_d$ (the result of
step 4) to the spectra in primary energy per nucleon. For each nucleus
this procedure includes firstly a calculation of expected primary
energy for each edge of the region of $E_d$ (see step 1). To solve
this problem FLUKA simulation of energy deposition was used with
supposition of the primary differential momentum spectra to be a power
law with index $\gamma = -2.6$ (there is only a weak dependence of
exact value of $\gamma$). After normalization of
the primary energy to the atomic weight and to the width of the
related energy region, the fluxes of different nuclei have
different energy per nucleon binning. To obtain the ratio
of fluxes at the same energy we calculate the energy points obtained
as a geometrical mean for the corresponding points for boron and
carbon and calculate all fluxes for these energy points by
interpolation of the spectrum of each nucleus. This procedure
increases B/C ratio from step 4 by 13\%--23\% (different for
different energies). 

6. The mean altitude of the flight of ATIC-2 was 36.5~km which
corresponds to 4.87~g/cm$^2$ of residual atmosphere. To obtain the
primary fluxes above the atmosphere, the interaction of
nuclei in the atmosphere should be accounted for. The interaction may
be described as fragmentation of nuclei without changing the energy
per nucleon. Then the interaction for each primary energy and for each
primary nucleus $q$ may be described by a set of coefficients
$L^q_{q'}, q' \le q$ which show the probability to find the nucleus
$q'$ at the entrance of the instrument. The coefficients $L^q_{q'}$ were
calculated by simulation of propagation of nuclei in the atmosphere by
FLUKA. Let $\psi_q$ be the flux of nucleus $q$ in 
energy per nucleon at the entrance of ATIC (these values are
known after step 5; we omit the index of the energy for simplicity)
and $\varphi_q$ be the same values above the atmosphere for some
definite energy per nucleon. Then one can write
\begin{equation}
  \label{eq:Phi}
  \begin{array}{l}
    \psi_4 = L^4_4\varphi_4 +\dots+L^{14}_4\varphi_{14} + \varepsilon_4\\
    \dots\dots\dots\dots\dots\dots\dots\dots\\
    \psi_{14} = L^{14}_{14}\varphi_{14} + \varepsilon_{14},
  \end{array}
\end{equation}
where $\varepsilon_4,\dots,\varepsilon_{14}$ are small corrections
related to the fragmentation of nuclei heavier than silicon $(q=14)$. 
If $\varepsilon_q$ are known then the system (\ref{eq:Phi}) is a
square linear system relative to $\varphi_4,\dots,\varphi_{14}$ with
the triangle and diagonally-dominated matrix $L^q_{q'}$ and it can be
solved directly starting from the final equation. The atmospheric
correction reduces B/C ratio by 13\%--33\% for different primary
energies. 

The experimental errors were calculated as a combination of the
Poisson dispersion of the experimental statistics and the statistical
errors of the simulations.  If the desired quantities were obtained as
a solution of a linear system (as in steps 4 and 6) then corresponding
complete covariation matrix were calculated by the Monte Carlo method.
All reported errors are standard deviations.

\section{Results and discussion}
\fussy

\begin{figure}[bt]
\begin{center}
\includegraphics [width=0.48\textwidth,height=0.32\textwidth]{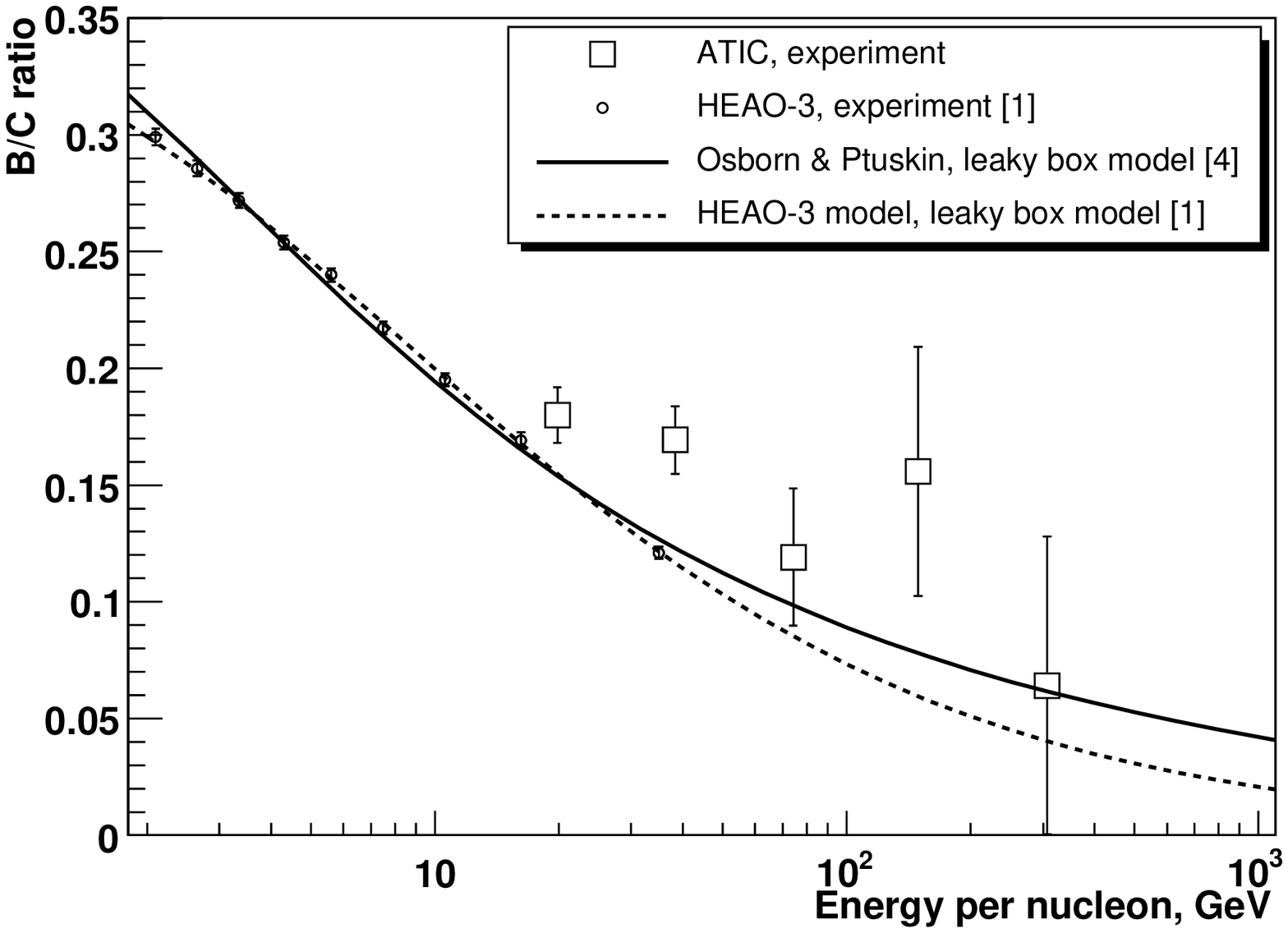}\\
\includegraphics [width=0.48\textwidth,height=0.32\textwidth]{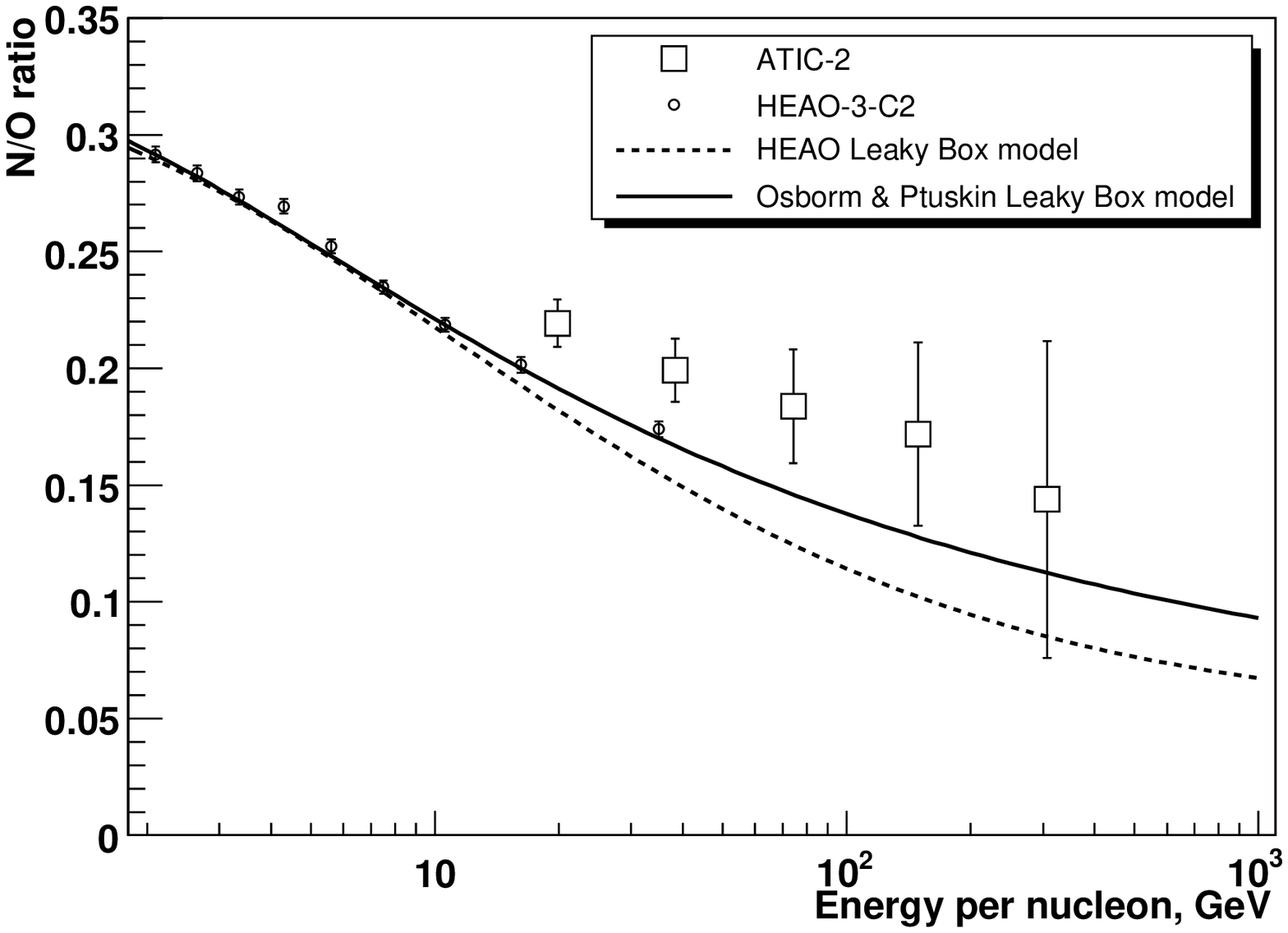}\\
\includegraphics [width=0.48\textwidth,height=0.32\textwidth]{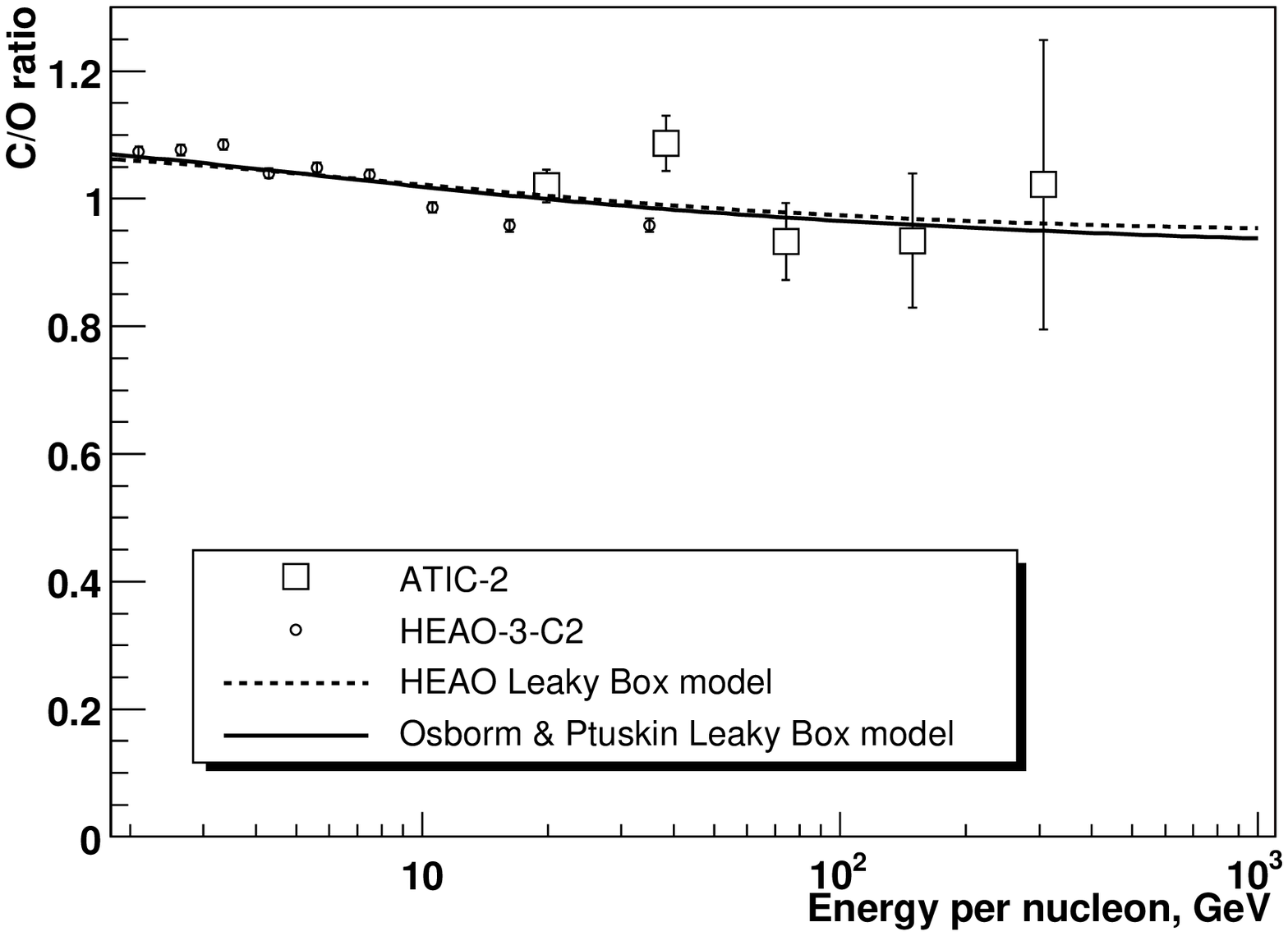}\\
\end{center}
\caption{B/C, N/O and C/O (top-down) ratio from this 
work, from HEAO-3-C2 experiment and from leaky box models.} 
\label{fig2}
\end{figure}

The results for B/C, N/O, C/O ratio are presented in table \ref{tab:1}. 
\begin{table}[t]
  \centering
  \caption{B/C, N/O, C/O ratios as a function of primary energy 
    (GeV/n). The numbers in parenthesis give the uncertainty in the 
    last significant digits quoted.} 
  \begin{tabular}{|c|c|c|c|}
  \hline
  $E$ & B/C &  N/O & C/O\\
  \hline
  19.9 & 0.180(11) & 0.219(10) & 1.020(26)\\
  38.3 & 0.169(15) & 0.199(13) & 1.087(43)\\
  74.3 & 0.119(29) & 0.184(24) & 0.933(60)\\
  149  & 0.156(53) & 0.172(39) & 0.934(105)\\
  307  & 0.064(63) & 0.144(68) & 1.022(227)\\
  \hline
  \end{tabular}
  \label{tab:1}
\end{table}
The data for B/C, N/O and C/O ratio along with the data of HEAO-3-C2
experiment \cite{HEAO_ENGELMANN1990} with theoretical predictions are
shown in fig.~\ref{fig2}. One can see that the data of present work
for B/C and N/O is somewhat above the data of
\cite{HEAO_ENGELMANN1990} but is extended to higher energies. The
theoretical curves in fig.~\ref{fig2} are calculations in leaky box
approximation. The dashed line is based on the HEAO-3-C2 fit for the
Galaxy escape length \cite{HEAO_ENGELMANN1990}
$\lambda_{\mbox{\scriptsize esc}} = 34.1\beta R^{-0.60}$~g~cm$^{-2}$
(R is the rigidity) and the solid line is for the escape length
obtained in the model of Kolmogorov type of magnetic turbulence and
reacceleration during
propagation \cite{OSBORN_PTUSKIN1988}:\\
$\lambda_{\mbox{\scriptsize esc}} = 4.2(R/R_0)^{-1/3}
\left[1+(R/R_0)^{-2/3}\right]$~g~cm$^{-2}$, where $R_0 = 5.5$~GV.
Whereas the experimental data support general trend of decreasing B/C
and N/O ratio with energy, it is impossible to distinguish between
different models of propagation of particles due to the experimental
uncertainties.

\sloppy
It should be noted that our experimental data are model dependent
due to extensive simulation of the backgrounds by the FLUKA
code, but this could be improved by usage of additional
simulation codes. One can expect that the experimental separation between
different models of propagation would be possible with additional
experiments.  


\emph{Acknowledgements.} This work was supported by RFBR Grant
05-02-16222 in Russia and NASA Grants Nos. NNG04WC12G, NNG04WC10G,
NNG04WC06G in the USA.

\end{document}